\documentclass[aps,prd,showpacs,nofootinbib,floats,floatfix,preprintnumbers,groupedaddress,twocolumn]{revtex4-1}
\usepackage{graphicx,epsfig}
\usepackage{dcolumn}
\usepackage{bm}
\usepackage{latexsym}
\usepackage{color}
\usepackage{tabularx}
\usepackage{ulem}
\usepackage{hyperref}
\usepackage{float}
\usepackage{tabularx}
\usepackage{color}
\usepackage{comment}
\usepackage{physics}

\usepackage{tikz}
\usepackage{hyperref}
\usepackage{colortbl}
\usepackage{listings}
\usepackage{amsmath,amsfonts,amssymb}
\usepackage{fancyhdr}
\usepackage{hyperref}
\usepackage{natbib}
\hypersetup{
	colorlinks   = true, 
	urlcolor     = blue, 
	linkcolor    = blue, 
	citecolor    = red 
}

\hypersetup{
	colorlinks   = true, 
	urlcolor     = blue, 
	linkcolor    = blue, 
	citecolor   = red 
}

\begin{document}
\title{Horizon thermalization of Kerr black hole through local instability}
\author{Surojit Dalui}
\email{suroj176121013@iitg.ac.in}
\author{Bibhas Ranjan Majhi}
\email{bibhas.majhi@iitg.ac.in}
\affiliation{Department of Physics, Indian Institute of Technology Guwahati, Guwahati 781039, Assam, India}

\date{\today}

\begin{abstract}
The validity of our already proposed conjecture -- {\it horizon creates a local instability which acts as
 the source of the quantum temperature of black hole} -- is being tested here for Kerr black hole. Earlier this has been explicitly shown for spherically symmetric static black hole (SSS BH). The more realistic situation like Kerr spacetime, being stationary and axisymmetric, is a non-trivial example to analyze. We show that for a chargeless massless particle, the near horizon radial motion in Kerr spacetime, like SSS BH, can be locally unstable. The radial contribution in the corresponding Hamiltonian is $\sim xp$ kind, where $p$ is the canonical momentum and $x$ is its conjugate position of particle. Finally we show that the horizon thermalization can be explained through this Hamiltonian when one dose a semi-classical analysis. It again confirms that near horizon instability is liable for its own temperature and moreover generalizes the validity of our conjectured mechanism for the black hole horizon thermalization.
\end{abstract}

\maketitle

{\section{\label{Intro}Introduction}}
The works of Bekenstein \cite{Bekenstein:1973ur} and Hawking \cite{Hawking:1974rv,Hawking:1974sw} indicate that the black hole (BH) horizon is a thermodynamical object, and the horizon temperature is the observer-dependent quantity. Interestingly, the concept of temperature came out to be purely quantum mechanical. Since then, understanding the thermalisation of the horizon in the context of black holes has been one of the central interests among physicists. The concept of horizon thermalisation originally originates from the analogy between the laws of black holes and those of the usual thermodynamical systems \cite{Bardeen:1973gs}. However, eventually, this analogy fails to give a unified perspective about the underlying mechanism which furnish temperature to the horizon. 

Recent developments towards this goal have revealed some striking insights on the connection between the horizon thermalisation and the local instability in the near horizon region \cite{Maldacena:2005he,Betzios:2016yaq,Morita:2019bfr,Hegde:2018xub,Maitra:2019eix,Setare:2021gzm,Dalui:2019esx,Dalui:2020qpt,Majhi:2021bwo,Subramanyan:2020fmx} 
{\footnote{Near horizon calculation is sufficient to see particle emission from horizon has also been reported in \cite{Camblong:2020pme,Azizi:2020gff}.}}. 
Although these two features may seem different from each other, there lies an important unification between them. In many works \cite{Morita:2019bfr,Hegde:2018xub,Maitra:2019eix,Setare:2021gzm,Dalui:2019esx,Dalui:2020qpt,Majhi:2021bwo,Subramanyan:2020fmx}, it has been noted that their intimate relationship has the potential to become one of the leading candidates, which can shed some light on the exact mechanism of horizon thermalisation. In this direction, our earlier works \cite{Dalui:2018qqv,Dalui:2019umw,Dalui:2019esx,Dalui:2020qpt} have revealed that in the presence of static spherically symmetric black hole (SSS BH), an outgoing massless and chargeless particle experiences instability in the near-horizon region. We also obtained that in the quantum scale, this instability provides temperature into the system, which divulges a noticeable conjecture that \textit{horizon creates a local instability in its vicinity region which acts as the source of the quantum temperature of a black hole}. Within this analysis very recently it has been argued that this conjecture may be supported by the theory of eigenstate thermalization hypothesis \cite{Majhi:2021bwo}.

However, in those works \cite{Dalui:2019esx,Dalui:2020qpt}, the analysis has been restricted only to SSS BH case. Therefore, this conjecture needs to be tested in some realistic situation in order to be more robust.  Kerr black hole is one of the prime examples of this kind. Kerr black holes are stationary rotating black holes. Rotating black holes are formed in the gravitational collapse of a massive spinning star, or the collision of a collection of compact objects or stars, and realistic collisions have non-zero angular momentum. Therefore, it is expected that all black holes in nature are rotating ones. So, the Kerr solution has astrophysical relevance, and this study in the near-horizon region of Kerr spacetime will provide us with a much more realistic picture of the explanation of horizon thermalisation.

Like earlier \cite{Dalui:2019esx,Dalui:2020qpt}, we start our analysis with a massless and chargeless particle moving under the Kerr spacetime, where the coordinate system adopted here is a generalised version of Eddington-Finkelstein (EF) coordinates to the rotating case. The particle moves along the null normal to the $t=$ constant and $r=$ constant surface. We obtain that the motion of the particle along this null trajectory faces instability in the near-horizon region. We also found that for such particular motion of the particle, the observed instability in the near horizon region is an observer independent phenomena. The observer associated with the generalised EF coordinate system quantifies that the radial motion of the particle as $r\sim e^{\kappa t}$ and correspondingly the momentum is given by $p_r\sim e^{-\kappa t}$ where $r$ is the generalised EF radial coordinates and $t$ is the generalised EF time coordinates. $\kappa$ is the surface gravity of black hole. Whereas in the polar or in the azimuthal direction, the instability is not present.  Next, we obtain the corresponding Hamiltonian of the system, and it comes out that the radial contribution to the Hamiltonian is of $xp$ kind. After analysing the classical picture, we next proceed towards the quantum regime, where we observed the automatic appearance of thermality into the system as a result of this aforesaid instability. We mainly concentrate on a semi-classical approach, namely tunneling formalism \cite{Srinivasan:1998ty,Parikh:1999mf,Banerjee:2008cf,Banerjee:2008sn,Banerjee:2009wb,Majhi:2011yi} through black hole horizon. The whole analysis establishes the generality of our aforesaid conjecture and thereby demands a possible candidate to explore the underlying theory of horizon thermalization process of a black hole.  


{\section{\label{Null hypersurface} Defining outgoing path}}
We start our analysis with the aim of identifying the outgoing trajectory of a massless, chargeless particle in Kerr black hole. This will be the backbone for the main purpose of the present work. The analysis is being followed from Appendix D of \cite{Gourgoulhon:2005ng}.
Massless particle follows null-like trajectories and therefore the tangent to the path must be null-like.  In Boyer-Lindquist (BL) coordinates ($t_{BL},r,\theta,\phi_{BL}$), the metric for Kerr black hole is of the form:
\begin{eqnarray}
ds^2=&&-\left(1-\frac{2m r}{\rho^{2}}\right)dt_{BL}^{2}-\frac{4m a r \sin^{2}\theta}{\rho^{2}}dt_{BL}d\phi_{BL} + \frac{\rho^{2}}{\Delta}dr^{2}\nonumber\\
&&+\rho^{2}d\theta^{2}+\left(r^{2}+a^{2}+\frac{2m r a^{2}\sin^{2}\theta}{\rho^{2}}\right)\sin^{2}\theta~d\phi_{BL}^{2}~,
\label{Kerr metric in BL}
\end{eqnarray}
where $m$ and $a$ are mass and angular momentum per unit mass of the black hole, respectively. $\rho^2$ and $\Delta$ are given by $\rho^{2}=r^{2}+a^{2}\cos^{2}\theta$ and $\Delta=r^{2}-2m r+a^{2}$. 
For $a=0$ the metric reduces to the Schwarzschild metric. However, being a generalization of Schwarzschild coordinates to the rotating case, these coordinates are singular on the event horizon $\mathcal{H}$ which corresponds to $\Delta=0$. The location of the horizon is then given by 
\begin{eqnarray}
r_{H}=m+\sqrt{m^{2}-a^{2}}~.
\label{horizon}
\end{eqnarray}
The metric coefficients are independent of $t_{BL}$ and $\phi_{BL}$ and thereby respecting time translation invariance and axial symmetry. 

 Since BL coordinates are not regular at $r_H$, like our earlier analysis in \cite{Dalui:2020qpt}, we adopt here a generalization of Eddington-Finkelstein (EF) coordinates to the rotating case which are regular at the horizon. The spheroidal version of EF coordinates, denoted by $(t,r,\theta,\phi)$,  are related to the Boyer Lindquist coordinates $(t_{BL},r,\theta,\phi_{BL})$ by the following relations:
\begin{eqnarray}
dt&=&dt_{BL}+\frac{dr}{\frac{r^{2}+a^{2}}{2mr}-1}~,
\label{relation BL and EF time}
\\
d\phi&=&d\phi_{BL}+\frac{a~dr}{r^{2}-2mr+a^{2}}~.
\label{relation BL and EF phi}
\end{eqnarray}
The metric in these coordinates turns out to be 
\begin{eqnarray}
ds^{2}=&&-\left(1-\frac{2mr}{\rho^{2}}\right)dt^{2}+\frac{4mr}{\rho^{2}}dt~dr-\frac{4amr}{\rho^{2}}\sin^{2}\theta~dtd\phi\nonumber\\
&&+\left(1+\frac{2mr}{\rho^{2}}\right)dr^{2}-2a\sin^{2}\theta\left(1+\frac{2mr}{\rho^{2}}\right)drd\phi\nonumber\\
&&+\rho^{2}d\theta^{2}+\left(r^{2}+a^{2}+\frac{2a^{2}mr\sin^{2}\theta}{\rho^{2}}\right)\sin^{2}\theta d\phi^{2}~.
\label{kerr metric in EF}
\end{eqnarray} 
Note that all the metric coefficients of the above one  are all regular at $r=r_{H}$. 

Now in order to find the null vector which will be the tangent to our particle path we first find the unit normals to $t=$ constant and $r=$ constant surfaces for the metric (\ref{kerr metric in EF}). The outward unit timelike normal to $t = $ constant surface is given by (see \cite{Gourgoulhon:2005ng} for details)
\begin{eqnarray}
n^{a}=\left(\frac{1}{\rho}\sqrt{\rho^{2}+2mr},-\frac{2mr}{\rho\sqrt{\rho^{2}+2mr}},0,0\right)~,
\label{timelike unit normal contravariant}
\end{eqnarray}
and the corresponding covariant vector is
\begin{eqnarray}
n_{a}=\left(-\frac{\rho}{\sqrt{\rho^{2}+2mr}},0,0,0\right)~.
\label{timelike unit normal covariant}
\end{eqnarray}
On the other hand the outward unit spacelike normal to $r=$ constant turns out to be
\begin{eqnarray}
s_{a}&=&\left(0,\rho\sqrt{\frac{\rho^{2}+2mr}{A}},0,0\right)~;
\\
s^{a}&=&\left(0,\frac{1}{\rho}\sqrt{\frac{A}{\rho^{2}+2mr}},0,\frac{a}{\rho}\sqrt{\frac{\rho^{2}+2mr}{A}}\right)~,\label{spacelike unit normal}
\end{eqnarray} 
where $A=(r^{2}+a^{2})^{2}-(r^{2}-2mr+a^{2})a^{2}\sin^{2}\theta$.
We can obtain the null normal $\mathbf{l}$ to the null surface by $\mathbf{l}=N(\mathbf{n}+\mathbf{s})$, where 
\begin{eqnarray}
N=\frac{\rho}{\sqrt{\rho^{2}+2mr}}~,
\label{lapse function}
\end{eqnarray}
is the Lapse function corresponding to our metric (\ref{kerr metric in EF}) (see Appendix D of  \cite{Gourgoulhon:2005ng} for details). Then the null vector turns out to be as
\begin{eqnarray}
l^{a}=\left(1,\frac{\sqrt{A}-2mr}{\rho^{2}+2mr},0,\frac{a}{\sqrt{A}}\right)~.
\label{null normal vector}
\end{eqnarray}
It can be checked that on the horizon $\mathcal{H}$ the components reduces to 
\begin{eqnarray}
l^{a}\stackrel{\mathcal{H}}=\left(1,0,0,a/(2mr_{H})\right)~,
\label{null vector on the horizon}
\end{eqnarray} 
which is the global timelike Killing vector $\xi^a = \xi_{(t)}^a+\Omega_H\xi_{(\phi)}^a$ whose vanishing norm defines the horizon of the black hole.

We now consider our chargeless, massless particle to follow a trajectory whose tangent is defined by (\ref{null normal vector}). Then, the integral curves $x^{a}(\mu)=(t,r,\theta,\phi)$ of $l^{a}$, characterized by
\begin{eqnarray}
\frac{dx^{a}(\mu)}{d\mu}=l^{a}(x^i(\mu))~,
\label{integral curves}
\end{eqnarray}    
where $\mu$ is the parameter which fixes the particle position at a particular moment, leads to the outgoing trajectory of our massless particle. 

{\section{\label{radial instability}Near horizon behaviour of trajectory}}
The path of our test particle is given by the integral curves which are expressed in Eq. (\ref{integral curves}). We would like to investigate those trajectories and try to understand how they behave in the near horizon region of the Kerr black hole.
Since the components of the normal vector $l^{a}$ is given by (\ref{null normal vector}) and $x^{a}=(t,r,\theta,\phi)$, the time component of (\ref{integral curves}) yields
\begin{eqnarray}
\frac{dt}{d\mu}=1\Rightarrow \mu=t\label{time component}~.
\end{eqnarray}
Therefore in this analysis the coordinate time $t$ plays the role of the parameter which fixes the moment of particle position.

\subsection{radial direction}
Since $\mu = t $, the radial motion from (\ref{integral curves}) is determined by
\begin{eqnarray}
\frac{dr}{dt}&=&\frac{\sqrt{A}-2mr}{\rho^{2}+2mr} \equiv F(r,\theta)~.
\label{radial equation}
\end{eqnarray}
Here we denoted the right hand side expression as $F(r,\theta)$ for future convenience. 
The solution of the above equation will give us the information about the behaviour of the particle trajectory in the radial direction. However, our main interest is to scrutinize the nature of the trajectory in the vicinity of the Kerr horizon. Therefore, we have to check the behaviour of the solution of Eq. (\ref{radial equation}) taking into account the near horizon approximation. It has already been checked that $F(r_{H},\theta)=0$ (see Eq. (\ref{null vector on the horizon})). Using then the Taylor series expansion of $F(r,\theta)$ around $r=r_{H}$ and considering only the leading order terms we obtain:
\begin{eqnarray}
F(r,\theta)\simeq\kappa(r-r_{H})~,
\label{near horizon approx}
\end{eqnarray}
where $\kappa=F'(r= r_{H},\theta)$ (the prime denotes the partial derivative with respect to the radial coordinate $r$) is identified as the surface gravity of Kerr black hole (see Appendix D.4 of \cite{Gourgoulhon:2005ng}).
The value of $\kappa$ is given by
\begin{eqnarray}
\kappa=\frac{r_{H}-m}{2mr_{H}}=\frac{\sqrt{m^{2}-a^{2}}}{2m(m+\sqrt{m^{2}-a^{2}})}~.\label{kappa}
\end{eqnarray} 
Note that $\kappa$ is a constant (independent of spacetime coordinates) and therefore the leading order term in the expansion of $F(r,\theta)$ (see Eq. (\ref{near horizon approx})) is function of radial coordinate only. Since we are interested to this order, denote Eq. (\ref{near horizon approx}) as
$f(r)\simeq\kappa(r-r_{H})$.
It may be mentioned that $\kappa$ is the value of non-affinity parameter $\tilde\kappa$, calculated at the horizon, which is determined by the non-affinely parameterized null geodesic equation $l^a\nabla_al_b =\tilde\kappa l_b$ for our given (\ref{null normal vector}). 
Therefore keeping upto the relevant leading order ( i.e. $ \mathcal{O}(r-r_{H})$), in the near horizon region Eq. (\ref{radial equation}) reduces to
\begin{eqnarray}
\frac{dr}{dt}\simeq\kappa(r-r_{H})~.\label{radial eqn near horizon}
\end{eqnarray}
The solution of it is given by
\begin{eqnarray}
r-r_{H}=\frac{1}{\kappa}e^{\kappa t}~,
\label{radial eqn solution}
\end{eqnarray}
which has exponentially growing behaviour with respect to $t$.
In later discussion we will find that the corresponding radial momentum behaves as $p_r\sim e^{-\kappa t}$ and as we have limit $t\rightarrow -\infty$ for near horizon, $p_r$ diverges when the particle goes very near to the horizon.
This above solution indicates that as long as the particle resides in the vicinity of the horizon, it feels an instability. However, this equation also tells us that this mentioned instability is a ``local" feature, not a global one. It is only applicable in the near horizon region. Incidentally, this exact feature has been experimentally verified in a recent investigation \cite{Analogexpt} on an analog black hole made by photon chip.  Moreover it must be noted that the horizon corresponds to $t\to -\infty$ which is consistent to our result $r-r_H=(1/\kappa)e^{\kappa t}$. So within the range $-\infty\leq t<0$ we have $\kappa(r-r_H)<1$ and hence there is no breakdown of our approximation.

{\subsection{\label{angular behaviour}Angular direction}}
Next let us investigate the particle motion along the angular coordinates. From the obtained path of the test particle, given by the integral curve (\ref{integral curves}) of the null normal vector (\ref{null normal vector}) we obtain from the $\theta$ component of $l^{a}$
\begin{eqnarray}
\frac{d\theta}{dt}=0\Rightarrow \theta= {\textrm{constant}}~.
\label{theta component eqn}
\end{eqnarray} 
Therefore, the above equation suggests that the $\theta$ coordinate remains constant with respect to time along the trajectory of the particle; i.e. the particle does not have any motion along $\theta$-direction.  

The azimuthal component of (\ref{null normal vector}) leads us to
\begin{eqnarray}
\frac{d\phi}{dt}=\frac{a}{\sqrt{A}}~.\label{azimuthal component eqn}
\end{eqnarray}
Now, as our interest is in the near horizon region, therefore considering only the leading order term in the above equation we obtain
\begin{eqnarray}
\frac{d\phi}{dt}\simeq \frac{a}{2mr_{H}}=\Omega_{H}~.\label{phi component eqn}
\end{eqnarray}
The solution of the above equation is
\begin{eqnarray}
\phi=\Omega_{H} t~; 
\end{eqnarray} 
i.e. $\phi$ changes linearly with $t$ along the trajectory as long as near horizon particle motion is concerned.

The analysis in this section indicates that the motion of the particle in the near horizon regime in the radial direction incorporates a ``local'' instability, whereas such is absent in the angular directions. In the next section, for this particular motion, we will show the presence of such instability in a coordinate independent way.
{\section{\label{Covariant local instability} Local instability in geodesic congruence}}
We have already obtained that there is a local instability in the radial direction of the particle motion in the near horizon region. However, in the azimuthal direction the instability is not present.
 It is always important to give a description of this local instability in a covariant way. The main emphasis is to check whether this instability in the particle motion is an artefact of our chosen coordinate system or this near horizon instability has any coordinate independent explanation. Moreover, the instability is mainly measured by the exponential nature of separation between the two nearby geodesics. Therefore it is natural to analysis the geodesics deviation equation  for our chosen null vector, given by null Raychaudhuri equation, in the near horizon regime. Our main focus hence will be on the expansion parameter $\Theta$, which measures the nature of distance between the geodesics in congruence. Also we will see whether this has any connection to our earlier prediction on radial instability. 

The Raychaudhuri equation for null geodesics is \cite{Paddybook}
\begin{eqnarray}
\frac{d\Theta}{d\mu}=\tilde{\kappa}\Theta-\frac{1}{2}\Theta^{2}-\sigma_{ab}\sigma^{ab}+\omega_{ab}\omega^{ab}-R_{ab}l^{a}l^{b}~,
\label{null Raychaudhuri}
\end{eqnarray} 
where $\tilde{\kappa}$ is the non-affinity coefficient, $\sigma_{ab}$ is the shear parameter, $\omega_{ab}$ is the rotation parameter and $R_{ab}$ is the Ricci tensor.    
Since our aim is to study this equation in the near horizon region of Kerr spacetime, we will now investigate nature of each terms in the above equation. Note that the quantities are defined with respect to the null vector (\ref{null normal vector}) as the motion of our particle is along this vector. Therefore, it is time to examine each of the terms on the right hand side of this equation and try to analyze them. We obtain that although the value of the shear parameter $\sigma_{ab}$ does not vanish for any arbitrary value of $r$ but on the horizon $r=r_{H}$ it vanishes, i.e. $\sigma_{ab}(r_{H})=0$ (see \cite{Gourgoulhon:2005ng} for detailed discussion). Therefore, in the near horizon region the leading order term of the shear parameter is
\begin{eqnarray}
\sigma_{ab}(r)\simeq \sigma'_{ab}(r_{H})(r-r_{H})~,\label{sigma near horizon}
\end{eqnarray}  
where the higher order terms $(\mathcal{O}(r-r_{H})^{2})$ have been  neglected and the prime denotes the derivative with respect to $r$. We have also found that on the horizon $\Theta\stackrel{\mathcal{H}}=0$. Therefore, in the near horizon region the leading order term of $\Theta$ which contributes in the Raychaudhuri equation is
\begin{eqnarray}
\Theta(r)\simeq\Theta'(r_{H})(r-r_{H})~,
\end{eqnarray} 
where, as earlier, the higher order terms have been neglected. Whereas for the non-affinity parameter we have $\tilde{\kappa}=\kappa+\mathcal{O}(r-r_{H})$. We must also have the rotation parameter $\omega_{ab}=0$, as $l_{a}$ is hypersurface orthonormal. On the other hand, since Kerr BH is the solution of vacuum Einstein’s field equations, we have $R_{ab}=0$. Therefore, keeping only the leading order terms, i.e. $\mathcal{O}(r-r_{H})$ terms in the right hand side of Eq. (\ref{null Raychaudhuri}) one obtains
\begin{eqnarray}
\frac{d\Theta}{d\mu}\simeq\kappa\Theta~.
\end{eqnarray} 
Hence the expansion parameter, in the near horizon regime, at the leading order behaves as
\begin{eqnarray}
\Theta \simeq \kappa e^{\kappa\mu}~.
\label{expansion parameter near horizon}
\end{eqnarray}
So, the above expression suggests that in the near horizon the the geodesic congruence expand exponentially with increase of $\mu$. This is the signature of the presence of instability in the geodesic motion of the particle in the vicinity of the Kerr horizon. As $\Theta$ is a scalar quantity and its nature has been extracted from a covariant equation, this above analysis provides a covariant description of our aforesaid local instability as long as particle trajectory is defined by (\ref{null normal vector}).

We will now check whether this observation has any direct connection with our radial instability, given by (\ref{radial eqn solution}).
The expansion parameter $\Theta$ can be expressed in terms of $l^{a}$ and $\tilde{\kappa}(r)$ as
\begin{eqnarray}
\Theta = \nabla_{a}l^{a}-\tilde{\kappa}~.\label{expansion parameter in terms l^a and kappa}
\end{eqnarray}  
Now, let us take $\nabla_{a}l^{a}=M(r)$ and in the near horizon region we can expand $M(r)$ as $M(r)\simeq \kappa + M'(r_{H})(r-r_{H})$, where the prime denotes the derivative with respect to the radial coordinate (a detailed derivation is sketched in Appendix \ref{App2}). Similarly, we can expand $\tilde{\kappa}(r)$ in the near horizon as $\tilde{\kappa}(r)\simeq \kappa + \tilde{\kappa}'(r_{H})(r-r_{H})$. Therefore, putting these values into Eq. (\ref{expansion parameter in terms l^a and kappa}) we obtain the value of expansion parameter in the near horizon region as
\begin{eqnarray}
\Theta = S(r_{H})(r-r_{H})\label{theta near horizon}
\end{eqnarray}
where $S(r_{H})=\left[M'(r_{H})-\tilde{\kappa}'(r_{H})\right]$ and $S(r_{H})\neq 0$. Now, substituting this value of $\Theta$ (\ref{theta near horizon}) in the solution (\ref{expansion parameter near horizon}) yields $r-r_{H}\simeq (\kappa/S(r_{H}))e^{\kappa t}$. Hence, once again we found that there is instability in the radial direction of the particle motion and the nature of instability is similar in nature (\ref{radial eqn solution}). 
With this analysis we conclude that although the local instability is not a coordinate dependent feature, but the particular nature of radial behaviour (\ref{radial eqn solution}) is related to our chosen coordinate system.
{\section{\label{Hamiltonian}The Hamiltonian}}
Let us now proceed to construct the Hamiltonian of the particle which leads to the above observed features in near horizon particle motion. This will be explicitly used later for the semi-classical analysis of our system.

\subsection{From the knowledge of trajectory}\label{From the knowledge of trajectory}

We have the near horizon radial motion which is given by (\ref{radial eqn solution}). Use of the Hamilton's equation of motion $\dot{r}=\partial H/\partial p_{r}$ implies
\begin{eqnarray}
\frac{\partial H}{\partial p_{r}}=\kappa (r-r_{H})~.
\end{eqnarray}
Solution of this is given by 
\begin{eqnarray}
H=\kappa(r-r_{H})p_{r}+h_{1}(r,\theta,p_{\theta},\phi,p_{\phi})~.
\end{eqnarray}
where $h_{1}(r,\theta,p_{\theta},\phi,p_{\phi})$ is some arbitrary function.

The polar equation of motion of the particle is given by Eq. (\ref{theta component eqn}). Therefore, use of Hamilton's equation of motion $\dot{\theta}=\partial H/\partial p_{\theta}$ implies
\begin{eqnarray}
\frac{\partial H}{\partial p_{\theta}}=0~,
\end{eqnarray}
and the solution of this is given by 
\begin{eqnarray}
H=h_{2}(r,p_{r},\theta,\phi,p_{\phi})
\end{eqnarray}
where $h_{2}(r,p_{r},\theta,\phi,p_{\phi})$ is again an arbitrary function. 

The equation of motion along the azimuthal direction is given by Eq. (\ref{azimuthal component eqn}). Therefore, using again the Hamilton's equation of motion $\dot{\phi}=\partial H/\partial p_{\phi}$ we obtain  
\begin{eqnarray}
\frac{\partial H}{\partial p_{\phi}}=\Omega_{H}~,
\end{eqnarray}
and its solution is given by 
\begin{eqnarray}
H=\Omega_{H} p_{\phi}+h_{3}(r,p_{r},\theta,p_{\theta},\phi)~,
\end{eqnarray}
where $h_{3}(r,p_{r},\theta,p_{\theta},\phi)$ is again some arbitrary function. Therefore, the Hamiltonian of the system becomes
\begin{eqnarray}
H=\kappa(r-r_{H})p_{r}+\Omega_{H}p_{\phi}+h(r,\theta,\phi)~.
\end{eqnarray}
Now, the value of the arbitrary function $h(r)$ can be fixed by using the information that the corresponding Lagrangian must vanish as we are dealing with massless particle. The Lagrangian for the above Hamiltonian comes out to be
\begin{eqnarray}
L=\sum_{i} p_{i}\dot{q}^{i} - H =-h(r,\theta,\phi)~.
\end{eqnarray}
Therefore, in order to make it vanish for the massless particle, we must choose $h(r,\theta,\phi)=0$. Thus we obtain our desired Hamiltonian in the near horizon of the Kerr BH and it is given by
\begin{eqnarray}
H=\kappa(r-r_{H})p_{r}+\Omega_{H}p_{\phi}~.
\label{near horizon Hamiltonian}
\end{eqnarray}
 Interestingly in this case $H-\Omega_H p_\phi = \kappa (r-r_H)p_{r}$ has the form of $xp$, like the SSS black hole \cite{Dalui:2019esx,Dalui:2020qpt}, and this is valid only in the near horizon at the leading order in $x$. As the above Hamiltonian is related to the null vector (\ref{null normal vector}), which has been constructed from the {\it outward} timelike and spacelike unit normals, it is related to the {\it outgoing} trajectory of our particle. Interesting observation is that $xp$ type term is always related to the conserved quantity corresponding to global timelike Killing vector $\xi^a = \xi^a_{(t)}+\Omega_H\xi^a_{(\phi)}$ whose vanishing norm defines the location of the horizon.  For instance here the conserved quantity is given by $K= -\xi^ap_a = E-\Omega_H p_\phi$ as $p_t = -E$. It may be noted that the solutions of the equations of motion (for radial trajectory only) corresponding to (\ref{near horizon Hamiltonian}) are $r\sim e^{\kappa t}$ and $p_r\sim e^{-\kappa t}$. The near horizon limit is achieved by considering $t\rightarrow -\infty$ and interestingly in this limit $p_r$ diverges. Therefore when the particle is approaching towards the horizon its motion feels a local instability which reaffirms our earlier demand.

At the end, we aim to possibility of thermalization of horizon through tunneling of particles through the horizon. In this case Hamiltonians for both outgoing and ingoing are essential. The approach, adopted so far, yields only the outgoing Hamiltonian. In the next section that for ingoing trajectory will be deduced using the dispersion relation among the four-momentum of the particle.  
{\subsection{\label{Hamiltonian from dispersion relation}Using dispersion relation}}
Another way to find out our desired Hamiltonian is from the dispersion relation which relates the four-momentum of our particle. It not only gives the Hamiltonian structure of the outgoing particle, it also provides the information about the ingoing one. We start with the Kerr metric written in $(t,r,\theta,\phi)$ coordinates, i.e. Eq. (\ref{kerr metric in EF}). The metric (\ref{kerr metric in EF}) is clearly stationary and axisymmetric and hence there are two associated Killing vectors: $\xi_{(t)}^{a}=(1,0,0,0)$ and $\xi_{(\phi)}^{a}=(0,0,0,1)$. Therefore we have two corresponding conserved quantities for a particle motion: energy $E=-\xi_{(t)}^{a}p_{a}=-p_{t}$ and the angular momentum $L_z= \xi_{(\phi)}^{a}p_{a}=p_{\phi}$ along the rotational axis of the black hole, where $p_{a}$ is the four momentum whose components are $p_{a}=(p_{t},p_{r},p_{\theta},p_{\phi})$. Now, using the covariant form of the dispersion relation $g^{ab}p_{a}p_{b}=0$ for the massless particle, we obtain the equation of the energy in terms of the radial and azimuthal components of momenta as
\begin{eqnarray}
g^{tt}E^{2}-2g^{tr}E p_{r}+2g^{r\phi}p_{r}p_{\phi}+&&g^{rr}p_{r}^{2}\nonumber\\
&&+g^{\phi\phi}p_{\phi}^{2}=0~,
\end{eqnarray}
where we have substituted $p_t = - E$ and since our particle has no motion along $\theta$ (see e.g.  Eq. (\ref{null normal vector}) we have used $p_\theta = 0$.
The above one yields two solutions of energy:
\begin{eqnarray}
E=\frac{g^{tr}}{g^{tt}}p_{r}\pm\Bigg[\left(\left(\frac{g^{tr}}{g^{tt}}\right)^{2}-\frac{g^{rr}}{g^{tt}}\right)p_{r}^{2}&&-2\frac{g^{r\phi}}{g^{tt}}p_{r}p_{\phi}\nonumber\\&&-\frac{g^{\phi\phi}}{g^{tt}}p_{\phi}^{2}\Bigg]^{\frac{1}{2}}~,
\label{energy solution}
\end{eqnarray}
where the values of each $g^{ab}$ components are given in Appendix \ref{App1}. Among these two solutions one corresponds to the energy of the outgoing particle and the other solution corresponds to the ingoing one. 

Let us now proceed to identify them. First investigate with the negative sign solution of Eq. (\ref{energy solution}). Note that the coefficient of $p_{r}$ in the first term of Eq. (\ref{energy solution}) is always a negative quantity for the particle outside the horizon (see the values of $g^{ab}$ components in the Appendix \ref{App1} (check Eq. (\ref{inverse metric})) whereas the second term ( the quantity with the square root) is a positive. On the other hand, since the energy $E$ is positive, the radial momentum $p_{r}$ is negative in this case. So, it suggests that the energy solution of Eq. (\ref{energy solution}) with the negative sign corresponds to the ingoing particle. Now, putting all the values of $g^{ab}$ components in this solution of energy and considering only the leading order terms in the near horizon region we obtain the energy for the ingoing particle as
\begin{eqnarray}
E=-R_{H}p_{r}-\Omega_{H}p_{\phi}+C_{H}~,
\label{BRM1}
\end{eqnarray}        
where $R_{H}=4mr_{H}/(\rho_{H}^{2}+2mr_{H})$ and $\rho_{H}=\rho(r_{H})$ and the value of $C_{H}$, which is unimportant for our present analysis, is given in the Appendix \ref{App1} ( see Eq. (\ref{CH})). From the above expression it is again evident that $p_r$ is negative and thereby confirming again that this energy is for ingoing particle.

Next, we shall investigate the other solution of Eq. (\ref{energy solution}), i.e. the energy value with the positive sign. As we have already obtained the energy solution with the negative sign corresponds to the ingoing particle, therefore consequently we can tell that the energy solution with the positive sign corresponds to the outgoing particle. However, in this case we want to affirm that this energy solution of the outgoing particle exactly matches with the energy solution which we obtained earlier (\ref{near horizon Hamiltonian}) with the knowledge of the outgoing particle trajectory (\ref{radial equation}). From the positive energy solution of (\ref{energy solution}), using Hamilton's equation of motion we obtain the radial equation of motion of the outgoing particle which is    
\begin{eqnarray}
\dot{r}=\frac{g^{tr}}{g^{tt}}+ \frac{\Bigg[\left(\left(\frac{g^{tr}}{g^{tt}}\right)^{2}-\frac{g^{rr}}{g^{tt}}\right)p_{r}-\frac{g^{r\phi}}{g^{tt}}p_{\phi}\Bigg]}{E-\frac{g^{tr}}{g^{tt}}p_{r}}~.\label{radial in terms of energy}
\end{eqnarray}  
Now for our already obtained outgoing radial path, given by (\ref{radial equation}), we want check whether the above yields the energy of the particle as (\ref{near horizon Hamiltonian}). Expressing Eq. (\ref{radial equation}) in terms of the $g^{ab}$ one finds
\begin{eqnarray}
\dot{r}=\frac{g^{tr}}{g^{tt}}+\sqrt{\left(\frac{g^{tr}}{g^{tt}}\right)^{2}-\frac{g^{rr}}{g^{tt}}}~.\label{radial eqn in terms gab}
\end{eqnarray}
Substituting the above expression in Eq. (\ref{radial in terms of energy}) and solving for $E$ we obtain
\begin{eqnarray}
&E=\left[\frac{g^{tr}}{g^{tt}}+\sqrt{\left(\frac{g^{tr}}{g^{tt}}\right)^{2}-\frac{g^{rr}}{g^{tt}}}\right]p_{r}-\frac{\frac{g^{r\phi}}{g^{tt}}}{\sqrt{\left(\frac{g^{tr}}{g^{tt}}\right)^{2}-\frac{g^{rr}}{g^{tt}}}}p_{\phi}~.
\label{energy soln in terms of gab}
\end{eqnarray}
Now, using the expressions of every $g^{ab}$ component in Eq. (\ref{energy soln in terms of gab}) and taking only the leading order term in the near horizon region we obtain the energy as (\ref{near horizon Hamiltonian}).
Therefore, we confirm here that the energy solution with positive sign of Eq. (\ref{energy soln in terms of gab}) corresponds to the energy of the outgoing particle.

{\section{\label{tunneling}Thermalization: a semi-classical approach}}

Through the tunneling approach we will now show that these Hamiltonians are responsible to feel the black hole horizon by the particle as thermal object.
In this approach the main quantity to be calculated is the tunneling probability which is a ratio between the outgoing and ingoing probability of particle through the horizon (see \cite{Srinivasan:1998ty} for the details).
We start with the standard ansatz for the wave function for a particle as
\begin{eqnarray}
\Psi(q^{\mu})=\exp\left[-\frac{i}{\hbar}S(q^{\mu})\right]~,
\label{ansatz wave function}
\end{eqnarray}
where $S(q^{\mu})$ is the Hamilton-Jacobi action for the particle which is defined as an integration of the momentum components $p^{\mu}$ of the particle with respect to the position coordinates $q^{\mu}$:
\begin{eqnarray}
S(q^{\mu})=\int p_{\mu}dq^{\mu}~.
\label{action gen form}
\end{eqnarray}
The outgoing and ingoing trajectories correspond to $\partial S/\partial q^{\mu}<0$ and $\partial S/\partial q^{\mu}>0$, respectively. Here we are interested to calculate the emission probability of the outgoing particle while the absorption probability of the ingoing one. The ratio of them will give our required tunneling probability.

First we calculate the emission action of the outgoing particle. The radial momentum is given by (\ref{near horizon Hamiltonian}). As the particle crosses the horizon from just inside to just outside, the integration limit must be chosen $x=-\epsilon$ to $x=\epsilon$ where $x\equiv r-r_{H}$ and $\epsilon>0$ and thus the ``emission" action is given by
\begin{eqnarray}
S[\text{Emission}]&=&\left(\frac{E-\Omega_{H}p_{\phi}}{\kappa}\right)\int_{-\epsilon}^{\epsilon}\frac{dx}{x}+\int_{0}^{2\pi} p_{\phi}d\phi
\nonumber
\\
&=&\left(\frac{E-\Omega_{H}p_{\phi}}{\kappa}\right)(-i\pi)+\text{Real part}~.
\end{eqnarray}
Note that for the above one $\partial S/\partial x = (E-\Omega_Hp_\phi)/(\kappa x)$ and as inside the horizon $x<0$, we have $\partial S/\partial x<0$. Therefore satisfies the outgoing trajectory condition as stated below Eq. (\ref{action gen form}).
The ``absorption" action for the ingoing particle can be calculated from radial momentum given by (\ref{BRM1}). Since it does not contain any singularity at $x=0$, one obtains
$S[\text{Absorption}]
=\text{Real quantity}~.$
Since for the tunneling probability, the explicit form of the above one is not necessary as long as the final value is real, we do not bother about its actual expression.
Therefore, the probability of emission turns out to be
\begin{eqnarray}
P[\text{emission}]&\sim & \Big|e^{-\frac{i}{\hbar}}S[\text{emission}]\Big|^{2}
\nonumber
\\
&\varpropto& \exp\left[-\frac{2\pi(E-\Omega_{H}p_{\phi})}{\hbar\kappa}\right]~,
\end{eqnarray}
whereas the probability of absorption is $P[\text{Absorption}]\sim 1$. Hence the tunneling probability is evaluated as
\begin{eqnarray}
\Gamma=\frac{P[\text{Emission}]}{P[\text{Absorption}]}\sim\exp\left[-\frac{2\pi(E-\Omega_{H}p_{\phi})}{\hbar\kappa}\right]~.\label{tunneling probability}
\end{eqnarray}
Hence use of the usual argument yields the expression for temperature as 
\begin{equation}
T = \frac{\hbar\kappa}{2\pi}~,
\label{BRM2}
\end{equation}
which is the well known Hawking expression \cite{Hawking:1974rv,Hawking:1974sw} for black hole horizon temperature.

Here we saw that the emission probability, calculated from the Hamiltonian (\ref{near horizon Hamiltonian}), is the main for showing the thermality of the horizon. We know that, this Hamiltonian, in the locality of horizon yields an instability in the radial motion of the particle. This is in favour of our earlier conjecture \cite{Dalui:2019esx,Dalui:2020qpt}, even for more general like axisymmetric black hole, that horizon creates a local instability in its vicinity which is responsible for its own thermalization as feels by our massless particle. 

In the present context we have shown the appearance of the horizon thermality in the tunneling approach only. There are other approaches where a direct quantum analysis of $xp$ Hamiltonian has been done. This also led to the thermalization of the horizon \cite{Dalui:2019esx,Dalui:2020qpt}. Also in literature it has been mentioned that $xp$ Hamiltonian (or its inverse harmonic oscillator form) inherently captures thermal character at the quantum level (e.g. see \cite{Maldacena:2005he,Betzios:2016yaq,Morita:2019bfr,Hegde:2018xub}). Since this Hamiltonian captures instability as well as thermal property and appeared here due to presence of horizon (which are being rigorously shown here), it is then clear that this near horizon instability may be liable for horizon thermalization.  

Let us now discuss the reason for proposing such a statement in this context elaborately. In the present scenario, we constructed a model where a probed massless and chargeless particle resides in the near-horizon region of a Kerr BH. The specific Hamiltonian for the outgoing particle we obtain in this case is of $xp$ kind in addition with some rotation part (see Eq. (\ref{near horizon Hamiltonian})). We observed that such a Hamiltonian provides a particle non-zero quantum probability of crossing the horizon. This is the main finding in this analysis to infer that the horizon has a temperature.  Now, the noticeable fact about this Hamiltonian is that it can be casted that of an inverted harmonic oscillator (IHO) in a new set of canonical variables $(X,P)$: $x=\frac{1}{\sqrt{2}}(P-X)$ and $p=\frac{1}{\sqrt{2}}(P+X)$ \cite{Book1}.  Furthermore, the IHO potential is unstable in nature. Therefore the test particle here experiences a ``local instability'' due to the presence of a horizon. This can be realised through the near horizon equations of motion as well. As discussed below Eq. (\ref{radial eqn solution}), the radial momentum diverges as one approached towards the horizon; i.e. $p_r\to\infty$ as $t\to -\infty$. Also note that $p_r$ diverges at the horizon for a given value of the conserved quantity $H-\Omega p_\phi$ (see Eq. (\ref{near horizon Hamiltonian})). Due to this particular instability at $r=r_H$, we found that the outgoing particle, at the quantum level, tunnels through a complex path. 
Therefore such singularity, on the one hand, provides instability in the particle motion and, on the other hand, plays a pivotal role in the calculation of the tunneling probability (\ref{tunneling probability}). We obtain a finite amount of tunneling probability of the particle for crossing the horizon with the correct classical limit (outgoing probability vanishes at $\hbar\to 0$). Moreover, such exponential behaviour of the expression is solely due to $xp$ structure of the Hamiltonian. In general, the time-reversal invariance dictates that the emission probability at a certain time is equal to the absorption probability for the same time and vice versa. Whereas our result in the present context is not consistent with this. It appears that the probability of emitting out of the particle through the horizon at a certain time is not equal to the probability of absorption into the horizon, resulting in an imbalance in the particle number between the outside and inside the region of the horizon. Furthermore, the exponential nature of the tunneling probability exhibits that the system is thermal in nature. The sole reason for thermality to show up in this calculation is due to the presence of this peculiar singularity at the horizon ($r=r_{H}$), which originates due to the $xp$ structure of the conserved quantity in the near horizon region. Therefore, it leads us to propose the conjecture that the near horizon local instability shows its presence at the quantum level through the horizon temperature, thereby perceiving the black hole as a thermal object.	

\section{\label{conclusion}Conclusion}
For the past few decades, physicists have been looking for a unified explanation to understand the underlying reason why the horizon is associated with temperature. In this scenario, recent developments \cite{Dalui:2019esx,Dalui:2020qpt,Majhi:2021bwo} have shown that the possible answer may be hiding behind the classical instability in the near horizon region and its intimate relationship with its quantum consequences. 


 However, in our earlier works \cite{Dalui:2019esx,Dalui:2020qpt}, this conjecture has been proposed by investigating only SSS BH scenario. Therefore, in order to attest to the validity of our conjecture, we needed to test it for a more generic class of black holes. Kerr black hole is one of the prime examples of that which is much more realistic than the SSS BH, and have astrophysical relevance. All known stars rotate, and realistic collisions have non-zero angular momentum, therefore, the black holes formed due to the collapse or collision of stars or compact objects are expected to be the rotating ones in nature. Hence, the foremost aim of this work was to test the robustness of the conjecture in a much more realistic picture in order to explain the horizon thermalization.   

Like our earlier investigation \cite{Dalui:2019esx,Dalui:2020qpt}, the main focus, in this paper, was to inspect whether the presence of local instability in the near horizon region is responsible for providing the temperature for a particular set of observer but this time in Kerr spacetime. We finally found that this is indeed the case. Although near horizon instability is an observer independent phenomenon for our chosen null path, the specific feature of the radial motion of the outgoing particle is a coordinate dependent event. The radial contribution in corresponding Hamiltonian of the system comes out to be of $xp$ kind, which is indeed an unstable Hamiltonian. This is where the classical part of the calculation ends. In the quantum regime, using the tunneling approach we obtain that thermality emerges due to that instability and the temperature exactly matches with the Hawking temperature. Therefore, it turns out that the extension of our proposed conjecture is applicable to much more generic black holes also. Hence, the generality of this conjecture evident here, and thereby providing a prospective to become one of the leading candidates to understand the underlying mystery of the horizon thermalization in the context of black holes.



\clearpage

\begin{widetext}
\begin{center}
{\bf Supplementary material}
\end{center}
\appendix
\section{{\label{App2}}Calculation of $\nabla_{a}l^{a}=M(r)$ in the near horizon region }
The null vector is given by (\ref{null normal vector}) along which our test particle is moving. Now, expanding $M(r)$ we find
\begin{eqnarray}
M(r)=\nabla_{a}l^{a}=\frac{1}{\sqrt{-g}}\partial_{r}\Big(\sqrt{-g}~l^{r}\Big)~.
\end{eqnarray}
Putting the value of $l^{r}$ and $g$ in the above equation and performing the partial derivative with respect to $r$ we obtain
\begin{eqnarray}
M(r)=\partial_{r}f(r)+\frac{2r}{\rho^{2}}f(r)~,
\end{eqnarray}
where $f(r)$ is already defined earlier, i.e. $f(r)=(\sqrt{A}-2mr)/(\rho^{2}+2mr)$ and $f(r_{H})=0$. Now, using the Taylor series expansion of the function $M(r)$ in the near horizon region we obtain
\begin{eqnarray}
M(r)\simeq M(r_{H})+ M'(r_{H})(r-r_{H})~,
\end{eqnarray}
where $M(r_{H})=\kappa$ and $M'(r_{H})=\left(\partial^{2}_{r}f(r)\right)_{r=r_{H}} + \frac{2r_{H}}{\rho_{H}^{2}}\kappa$ with
\begin{eqnarray}
\left(\partial^{2}_{r}f(r)\right)_{r=r_{H}}=\frac{1}{2(\rho_{H}^{2}+2mr_{H})}\Bigg[\frac{1}{2mr_{H}}\left(12r_{H}^{2}+4a^{2}-2a^{2}\sin^{2}\theta\right) -&& \frac{1}{16m^{3}r_{H}^{3}}\left(4r_{H}(r_{H}^{2}+a^{2}) - 2(r_{H}-m)a\sin^{2}\theta\right)\nonumber\\
&&-\frac{4\kappa(r_{H}+m)}{\rho_{H}^{2}+2mr_{H}}\Bigg]~.
\end{eqnarray}

\section{{\label{App1}}Inverse of metric (\ref{kerr metric in EF}) and the value of $C_{H}$ in Eq. (\ref{BRM1})}
Our massless and chargeless test particle is moving under the background of the metric given by (\ref{kerr metric in EF}). Now the determinant of the metric components (\ref{kerr metric in EF}) takes a very simple form
\begin{eqnarray}
\text{det}(g_{ab})=-\rho^{4}\sin^{2}\theta~,
\label{determinant of the metric}
\end{eqnarray}  
and the inverse metric takes the form
\begin{eqnarray}
g^{ab}=
\begin{pmatrix}
-\left(1+\frac{2mr}{\rho^{2}}\right) & \frac{2mr}{\rho^{2}} & 0 & 0\\
\frac{2mr}{\rho^{2}} & \frac{r^{2}+a^{2}-2mr}{\rho^{2}} & 0 & \frac{a}{\rho^{2}}\\
0 & 0 & \frac{1}{\rho^{2}} & 0\\
0 & \frac{a}{\rho^{2}} & 0 & \frac{1}{\rho^{2}\sin^{2}\theta}
\end{pmatrix}~.
\label{inverse metric}
\end{eqnarray}
Now, let us denote $X^{2}(r)=\left(\frac{g^{tr}}{g^{tt}}\right)^{2}-\frac{g^{rr}}{g^{tt}}$, $Y(r)=\frac{g^{r\phi}}{g^{tt}}$ and $Z(r)=\frac{g^{\phi\phi}}{g^{tt}}$. Therefore, in terms of these, the energy solution with the negative sign of Eq. (\ref{energy solution}) can be written as
\begin{eqnarray}
E&=&\frac{g^{tr}}{g^{tt}}p_{r}-\left[X^{2}(r)p_{r}^{2}-2Y(r)p_{r}p_{\phi}-Z(r)p_{\phi}^{2}\right]^{\frac{1}{2}}\nonumber\\
&=&\frac{g^{tr}}{g^{tt}}p_{r}-\left(X(r)p_{r}-\frac{Y(r)}{X(r)}p_{\phi}\right)
\left[1-\frac{\left(\frac{Y^{2}(r)}{X^{2}(r)}+Z(r)\right)p_{\phi}^{2}}{\left(X(r)p_{r}-\frac{Y(r)}{X(r)}p_{\phi}\right)^{2}}\right]^{\frac{1}{2}}\label{ingoing energy}~.
\end{eqnarray} 	
Now, performing the Binomial series expansion of the term under the square root of the above equation we obtain
\begin{eqnarray}
E=\left(\frac{g^{tr}}{g^{tt}} - X(r)\right)p_{r}&&+\frac{Y(r)}{X(r)}p_{\phi}\nonumber\\
&&+\underbrace{\left(X(r)p_{r}-\frac{Y(r)}{X(r)}p_{\phi}\right)\left[\frac{1}{2}\frac{\left(\frac{Y^{2}(r)}{X^{2}(r)}+Z(r)\right)p_{\phi}^{2}}{\left(X(r)p_{r}-\frac{Y(r)}{X(r)}p_{\phi}\right)^{2}}+\frac{1}{8}\frac{\left(\frac{Y^{2}(r)}{X^{2}(r)}+Z(r)\right)^{2}p_{\phi}^{4}}{\left(X(r)p_{r}-\frac{Y(r)}{X(r)}p_{\phi}\right)^{4}}+\dots\right]}_{C(r)}~.\label{CH}
\end{eqnarray}
Therefore, considering only the leading order terms in the near horizon region and putting all the values of $g^{ab}$ components we land up to Eq. (\ref{BRM1}) where $C_{H}=C(r=r_{H})$.
\end{widetext}

	
\end{document}